\documentclass[aps, a4paper, superscriptaddress, nofootinbib, twocolumn, 10pt]{revtex4}
\usepackage{amsmath,amssymb,bm,natbib}
\usepackage[colorlinks,citecolor=blue,urlcolor=blue,linkcolor=blue]{hyperref}
\usepackage{graphicx,colordvi}
\usepackage{booktabs,tabulary}
\usepackage{dsfont}
\usepackage{color}
\usepackage{epstopdf}  
\usepackage{fix-cm}

\usepackage{ulem}
\newcommand{\be}{\begin{equation}}
\newcommand{\ee}{\end{equation}}
\newcommand{\beq}{\begin{equation}}
\newcommand{\eeq}{\end{equation}}
\newcommand{\bea}{\begin{eqnarray}}
\newcommand{\eea}{\end{eqnarray}}

\newcommand{\gev}{\, \text{GeV}}

\newcommand{\el}{\text{el}}
\newcommand{\inel}{\text{inel}}

\newcommand{\nuth}{\nu_{th}}

\newcommand{\alphaem}{\alpha_\text{em}}
\newcommand{\mN}{m}
\newcommand{\mpi}{m_\pi}

\renewcommand{\Im}{\text{Im}\,}
\newcommand{\thr}{\text{th}}

\newcommand{\bsp}{\begin{sloppypar}}
\newcommand{\esp}{\end{sloppypar}}

\newcommand{\al}{&\!}

\RequirePackage{graphicx}

\begin{document}

\title{Analyticity and Regge asymptotics in virtual Compton scattering on the nucleon}
\author{Irinel Caprini}
\affiliation{Horia Hulubei National Institute for Physics and Nuclear Engineering, POB MG-6, 077125 Bucharest-Magurele, Romania}

\begin{abstract}We test the consistency of the data on  the nucleon structure functions  with analyticity and the  Regge asymptotics of the virtual Compton amplitude.  By solving a functional extremal problem, we derive an optimal lower bound on the maximum difference between the exact amplitude  and the dominant Reggeon contribution for energies $\nu$ above a certain high value $\nu_h(Q^2)$. Considering in particular the  difference of the amplitudes $T_1^\inel(\nu, Q^2)$ for the proton and neutron, we find that the lower bound decreases in an impressive way when $\nu_h(Q^2)$ is increased, and represents a very small fraction of the magnitude of the dominant Reggeon.  While the method cannot rule out the hypothesis of a fixed Regge pole, the results indicate that the data on the structure function are  consistent with an asymptotic behaviour given by leading Reggeon contributions. We also show that the minimum of the lower bound as a function of the subtraction constant $S_1^\inel(Q^2)$ provides a reasonable estimate of this quantity, in a frame similar, but not identical to the Reggeon dominance hypothesis.
\end{abstract}
\maketitle

\section{Introduction}\label{sec:introduction}
The virtual Compton scattering on the nucleon, in particular the doubly-virtual forward scattering $\gamma^*(q)+N(p)\to \gamma^*(q)+N(p)$,
 is  of much interest for the calculation of the electromagnetic mass difference between the proton and the neutron by Cottingham formula \cite{Cottingham:1963zz} and for the study of nucleon polarizabilities.   This process has been intensively studied a long time ago  \cite{Harari:1966mu, Gross:1968zz,  Creutz:1968ds, Damashek:1969xj, Dominguez:1970wu, Elitzur:1970yy, Brodsky:1971zh, Zee:1971df, Brodsky:1972vv},  especially in the frame of Regge theory. In particular, a question that received much attention was the possible existence of a fixed  pole at $J=0$ in the angular momentum plane for the Compton amplitude. This problem was discussed also  more recently, see for instance \cite{Brodsky:2008qu, Muller:2015vha} and references therein. 

Dispersion theory is an important tool for investigating the virtual Compton amplitude.\footnote{Lattice calculations started to produce competitive results in recent years. For an updated comparison of lattice and dispersive results, see Fig. 8 of \cite{Gasser:2020hzn}.}.  
Early dispersive evaluations of Cottingham formula have ben performed in \cite{Harari:1966mu, Gross:1968zz,  Damashek:1969xj, Dominguez:1970wu, Elitzur:1970yy, Gasser:1974wd}. There has been a recent renewed interest in this problem:  dispersive treatments of the Compton amplitude, using modern data on the structure functions, have been reported in
 \cite{WalkerLoud:2012bg, WalkerLoud:2012en, Erben:2014hza, Thomas:2014dxa, Gasser:2015dwa, Caprini:2016wvy, Walker-Loud2018, Tomalak:2018dho,  Gasser:2020hzn, Gasser:2020mzy}.

The  forward virtual Compton scattering is described by the time-ordered product of  two electromagnetic currents between on-shell proton states, summed over the spin orientations: 
\be \label{eq:Compton} T^{\mu\nu}(p,q)=\text{\small $\frac{i}{2}$}\!\int\!\! d^4x\,e^{i q\cdot x}\langle p|Tj^\mu(x) j^\nu(0)|p\rangle.\ee
By Lorentz covariance, this tensor  is decomposed as
\bea \label{eq:TK} \al\al T^{\mu\nu}(p,q)= (q^\mu q^\nu-g^{\mu\nu}q^2) T_1(\nu,Q^2)+ \\
\al\al  \frac{1}{m^2}\{(p^\mu q^\nu+p^\nu q^\mu)p\cdot q
 -g^{\mu\nu}(p\cdot q)^2-p^\mu p^\nu q^2\}T_2(\nu,Q^2), \nonumber
\eea
in terms of two invariant amplitudes, $T_1(\nu, Q^2)$ and $T_2(\nu, Q^2)$, free of kinematical singularities and zeros.   We recall the notations:  $\nu=p\cdot q/m$ and $Q^2=-q^2$, where $p$ and $q$ are the nucleon and photon momenta and $m$ is the nucleon mass.

 The invariant amplitudes  are even functions of $\nu$ due to  crossing symmetry. Moreover, they
  are real analytic functions in the $\nu^2$ complex plane, with singularities along the real axis imposed by unitarity:  poles due to the elastic contributions, 
 and cuts produced by the inelastic states.  Schwarz reflection principle implies that the discontinuities across the cuts are related to the imaginary parts on the upper edge of the cut. By unitarity, the contribution of the inelastic states is: 
\beq
\Im T_i(\nu^2,Q^2)=\pi V_i(\nu,Q^2),\quad \nu\ge \nuth, \quad i=1,2, \label{eq:ImT}
\eeq
where  $\nuth= \mpi+(\mpi^2-Q^2)/2\mN$ is the lowest threshold due to the $\pi N$ intermediate state,  and for $Q^2\ge 0$ the structure functions $V_i$ are obtained from the cross sections of lepton-nucleon scattering.

Besides analyticity, the asymptotic behaviour of the amplitudes for large $\nu$ at fixed $Q^2$  plays an important role in the dispersion theory.
It is generally accepted that the amplitude $T_2(\nu, Q^2)$ can be expressed by an unsubtracted dispersion relation, while the dispersion relation for $T_1(\nu, Q^2)$ requires a subtraction. This means that for recovering the amplitude, the structure function is not sufficient, one needs also the subtraction function, i.e. the value of the amplitude at a certain point, for instance $S_1(Q^2)\equiv T_1(0, Q^2)$. It turns out that the subtraction function plays a crucial role in the  dispersion theory, in particular for the proton-neutron electromagnetic mass difference. As we will discuss below, there are several different approaches to the determination of the subtraction function, and they depend on the specific assumptions about the  behaviour of the amplitude at high energies.

 The asymptotic behaviour
of the Compton amplitude is assumed in general to be governed by Reggeon exchanges, which manifest as moving poles in the angular momentum plane.
For the proton and neutron amplitudes, the dominant contributions are due to the exchange of the Pomeron. A Reggeon with the quantum numbers $I^C = 1^+$ and a trajectory with intercept around 0.5, identified with $a_2$, is the most important non-leading contribution.

 In Ref.  \cite{Gasser:1974wd} it was assumed that the Reggeons dominate the asymptotic
behaviour. This assumption was implemented by the requirement
\beq\label{eq:RD}\lim_{\nu\rightarrow\infty}\, [T_1(\nu, Q^2)-T_1^{R}(\nu, Q^2)]= 0,
\eeq
where $T_1^R(\nu, Q^2)$ is the leading Regge contribution. This \textit{Reggeon dominance} hypothesis  implies that the amplitude $T_1(\nu, Q^2)$  does not contain a fixed pole at $J=0$, which would contribute with a real constant term at large $\nu$.

 As shown in \cite{Gasser:1974wd}, by exploiting (\ref{eq:RD}) one can derive a sum rule which uniquely determines the subtraction function $S_1(Q^2)$ in terms of the cross section of lepton-nucleon scattering and the residues of the Regge poles.  A variant of this sum rule was proposed by Elitzur and Harari \cite{Elitzur:1970yy}, on the basis of duality and finite energy sum rules. The  recent analyses with modern data on the structure functions, performed in \cite{Gasser:2015dwa, Gasser:2020hzn, Gasser:2020mzy}, show that this sum rule leads to precise dispersive determinations of proton-neutron electromagnetic mass difference and nucleon polarizabilities.  

We mention also a more general frame of exploiting Regge asymptotics, investigated in \cite{Caprini:2016wvy}, where it was assumed that the modulus of the amplitude is given  above a certain energy by the modulus of Reggeon exchanges. Using this assumption,  upper and lower bounds on the subtraction function have been derived,  which turned out to be consistent with the sum-rule predictions made in \cite{Gasser:2015dwa}.

On the other hand, in the analyses performed in \cite{WalkerLoud:2012bg, WalkerLoud:2012en, Erben:2014hza, Thomas:2014dxa, Walker-Loud2018},   the knowledge of the Regge behaviour of the amplitude at high energies was not taken into account in an explicit way. As a consequence, the subtraction function was assumed to be an independent quantity, which cannot be calculated in terms of the lepton-nucleon cross sections. Instead,  the function $S_1(Q^2)$ was parametrized  by simple algebraic expressions, which interpolate between the low and high $Q^2$ regimes, known from effective field theory and perturbative QCD, respectively.  A detailed comparison of the predictions of these models with the subtraction function derived from Reggeon dominance is presented in Sect. 24.2 of \cite{Gasser:2020hzn}.

From the above discussion, it follows that the assumptions about the validity of  Regge asymptotics in virtual Compton scattering play an important role in the dispersive predictions.  In the present paper, we propose a test of Regge asymptotics, which exploits analyticity and the phenomenological input on the nucleon structure functions, available up to a certain energy. The question that we consider is whether this information can say something about the difference between the exact amplitude and the Reggeon contributions above that energy. 

It turns out that the standard dispersion relations are not useful for providing an answer to this question. Fortunately, there exist also other methods of exploiting analyticity, based on functional analysis and optimization theory (for a review, see \cite{Caprini:2019osi}). By solving an extremal problem on a suitable class of analytic function, we  obtain a rigorous lower bound on the maximum  difference between the true amplitude and the leading Reggeon contributions above a certain  energy. As we will show in the paper, the value of this lower bound and its variation when the energy is increased offer insights on the onset of the Regge regime.  By solving also a more general extremal problem, we investigate the dependence of the lower bound on the subtraction function and argue that a reasonable prescription for finding this quantity assuming Regge asymptotics is to look for the value that achieves the minimum of the lower bound. 

The paper is organized as follows: in the next section we specify the frame and formulate the extremal problems of interest for our study. In Sect. \ref{sec:input} we briefly discuss the phenomenological input used in the calculations.  The results are presented in Sect. \ref{sec:results}, while Sect. \ref{sec:conclusions} contains a summary and our conclusions. Finally, in appendix  \ref{sec:solution} we present the solution of the extremal problems formulated in Sect. \ref{sec:extremal}.

 \section{Extremal problems}\label{sec:extremal}
To illustrate our approach we consider the difference of the invariant amplitudes $T_1$ for the proton and neutron. 
 Also, in order to facilitate the comparison with the previous works \cite{Gasser:2015dwa, Caprini:2016wvy},  we consider  only the ``inelastic'' amplitude: 
\beq\label{eq:inel}
T_1^\inel(\nu, Q^2)\equiv T_1(\nu, Q^2)-T_1^\el(\nu, Q^2),
\eeq
defined  by subtracting from the total amplitude the elastic contribution  due to the nucleon pole. The explicit expression of  $T_1^\el(\nu, Q^2)$ is given in  \cite{Gasser:2015dwa}.

As discussed above, the amplitude $T_1^\inel(\nu, Q^2)$  obeys at fixed $Q^2$  a subtracted dispersion relation in the $\nu$ variable. Choosing the subtraction point at $\nu=0$, the dispersion relation has the form
\beq\label{eq:DRinel}
 T_1^\inel(\nu,Q^2)= S_1^\inel(Q^2)+2\nu^2\int_{\nuth}^\infty \frac{d\nu'}{\nu'} \,\frac{V_1(\nu',Q^2)}{\nu'^2-\nu^2-i\epsilon},
\eeq
where $\nuth$ is defined below Eq. (\ref{eq:ImT}) and 
\be 
S_1^\inel(Q^2)\equiv T_1^\inel(0,Q^2).\label{eq:S1}
\ee
As already mentioned, the dispersion relation, which is based on the Cauchy integral relation, is not the only way to implement the analyticity of the amplitude in the $\nu$ complex plane.  In our approach, analyticity will be exploited by functional-analysis techniques. We first specify the ingredients of this approach. 

Let $\nu_h(Q^2)$ be a certain high energy, which will be treated as a free parameter in the analysis.
Below this energy,  we  assume that the imaginary part is known  on the unitarity cut, i.e. we have
\beq
\Im T_1^\inel(\nu,Q^2)=\pi V_1(\nu,Q^2),~~~~  \nuth\leq \nu < \nu_h(Q^2). \label{eq:ImTinel}
\eeq

 At high energies,  the amplitude is described by the the exchange of  Reggeons. Since we consider the difference of the  proton and neutron amplitudes,  the Pomeron contributions  cancel out and the asymptotic behaviour is given by the leading  Reggeon. We write this contribution as\footnote{A sligthly different representation, given in Eq. (18) of \cite{Gasser:2020hzn}, coincides with (\ref{eq:TR})  at high $\nu$.} 
\beq\label{eq:TR}
T_1^{R}(\nu, Q^2)=-\pi  \beta (Q^2)\frac{1+\exp[-i\pi\alpha]} {\sin\pi \alpha }\,\nu^\alpha,
\eeq
where $\alpha>0$ is the $a_2$ trajectory at zero momentum transfer. Subleading contributions are expected to be present at finite energies.

It is of interest to define  a  quantity which measures how much the exact amplitude deviates from the prediction of the Regge model for  $\nu\ge \nu_h(Q^2)$. Taking into account the fact that the amplitude exhibits asymptotically a growth like $\nu^\alpha$ with $\alpha>0$, a reasonable candidate is
\be\label{eq:diffnu}
\sup_{\nu \ge \nu_h(Q^2)}\left |\frac{T^{\inel}_1(\nu, Q^2) -T_1^R(\nu, Q^2)}{\nu^\alpha}\right |,
\ee
which involves the difference of functions bounded in the complex plane. 

Since the physical amplitude is not known for $\nu\ge \nu_h(Q^2)$, the exact value of this supremum defined in (\ref{eq:diffnu}) is not known.  However, it turns out that it must exceed a certain value, which can be calculated. Namely, let us consider the functional extremal problem
\be\label{eq:delta0R}
\delta_0^{R}\!=\!\min_{\{T_1^\inel\}} \left[\sup_{\nu \ge \nu_h}\left |\frac{T_1^\inel(\nu, Q^2) -T_1^{R}(\nu, Q^2)}{\nu^\alpha}\right |\right],
\ee
where the minimization is taken upon the class of functions $\{T_1^\inel\}$ which are real-analytic in the $\nu^2$ plane cut for $\nu^2\ge \nu_{\thr}^2$, obey the unitarity condition (\ref{eq:ImTinel}) and exhibit an asymptotic increase bounded by $\nu^\alpha$. Since the true amplitude belongs to this class, the difference 
(\ref{eq:diffnu}) corresponding to it must be larger than $\delta_0^R$. Therefore, $\delta_0^R$ is a lower bound on the exact difference (\ref{eq:diffnu}). Moreover, unlike the exact difference which is not known, the lower bound can be calculated in terms of the phenomenological input.

The above extremal problem can be generalized to the case when more informations  about the class of analytic functions $T_1^\inel$  are available. Suppose  that the value of $T^\inel(\nu, Q^2)$  at a point $\nu_1$ inside the analyticity domain is given at each $Q^2$. We take in particular $\nu_1=0$, because this was the choice of the subtraction point in the dispersion relation (\ref{eq:DRinel}), but the more general case of an arbitrary  $\nu_1<\nu_\thr$ can be treated in a similar way. Then, we consider the modified extremal problem
\beq\label{eq:delta0Rhat}
\widehat\delta^R_0(S_1^\inel)\!=\!\min_{\{ T_1^\inel\}} \left[\sup_{\nu \ge \nu_h}\left |\frac{T_1^\inel(\nu, Q^2) -T_1^{R}(\nu, Q^2)}{\nu^\alpha}\right |\right],
\eeq
where the minimization is performed upon the class of functions $\{T_1^\inel\}$ with the properties specified below (\ref{eq:delta0R}), which satisfy in addition the constraint (\ref{eq:S1}). Therefore,  $\widehat\delta^R_0(S_1^\inel)$ is a lower bound on the difference between the physical amplitude $T_1^\inel$ and the Regge expression $T_1^R$ for $\nu>\nu_h$, when the value of the physical amplitude at $\nu=0$ is supposed to be given.  The notation in  (\ref{eq:delta0Rhat}) emphasizes the fact that the lower bound depends on the input value $S_1^\inel(Q^2)$.

The solution of the extremal problems (\ref{eq:delta0R}) and (\ref{eq:delta0Rhat}) is presented in appendix \ref{sec:solution}), where it is shown that the lower bounds $\delta_0^R$ and $\widehat \delta_0^R$ are obtained by a relatively simple numerical algorithm. Before discussing the significance and the applications of these quantities, we briefly review in the next section the phenomenological input  used in the calculations.

\section{Phenomenological input} \label{sec:input}

We used as input modern data on the nucleon structure functions.\footnote{I am grateful to J. Gasser and H. Leutwyler for  detailed information on the experimental input.} Experimental measurements are available from various groups, depending on the c.m.s. energy $W$, defined as
\be\label{eq:W}
W=\sqrt{s}=\sqrt{m^2+2 m \nu -Q^2}.
\ee
At low energies, $W\leq 1.3 \gev$,    we used the data on the cross sections measured in \cite{Drechsel:2007if, Kamalov:2000en, Hilt:2013fda, MAID}. 
In the range $1.3<W \leq 3 \gev$ we used the parametrization of Bosted and Christy \cite{Bosted:2007xd, Christy:2007ve}. 
For $W> 3\gev$ and  $Q^2\leq 1 \gev^2$, the input has been taken from the vector-meson dominance model of Alwall and Ingelman \cite{Alwall:2004wk}. For  $W> 3\gev$ and  $1<Q^2 \leq 3 \gev^2$,  we used the solution of the DGLAP equations constructed by Alekhin, Bl\"umlein and Moch, who obtained numerical values for the structure functions over a wide range: $1 \leq Q^2\leq2\cdot 10^5 $ and $10^{-7}\leq x\equiv Q^2/(2m\nu)\leq 0.99$. The underlying analysis is described in \cite{Alekhin:2013nda, Alekhin:2017kpj, Alekhin:2019ntu}. As discussed in \cite{Gasser:2015dwa, Gasser:2020hzn}, a conservative attitude is to attach large overall uncertainties, of about $30\%$,  to the values of the structure function in this range.

The parametrization of the structure function for $W> 3\gev$ and $Q^2\leq 1 \gev^2$ proposed in \cite{Alwall:2004wk} tends in the limit of high $\nu$ to the imaginary part of the Regge expression (\ref{eq:TR}) for  $\alpha=0.55$, allowing us to extract the   $Q^2$-dependent  residue  $\beta(Q^2)$. For $Q^2>1 \gev^2$, the residue is obtained as in \cite{Gasser:2020hzn}    from a Regge fit of the numerical table given in \cite{Alekhin:2013nda}.   

\begin{figure}[ht]
\centering \vspace{0.2cm}
\includegraphics[width=\linewidth, width=7cm]{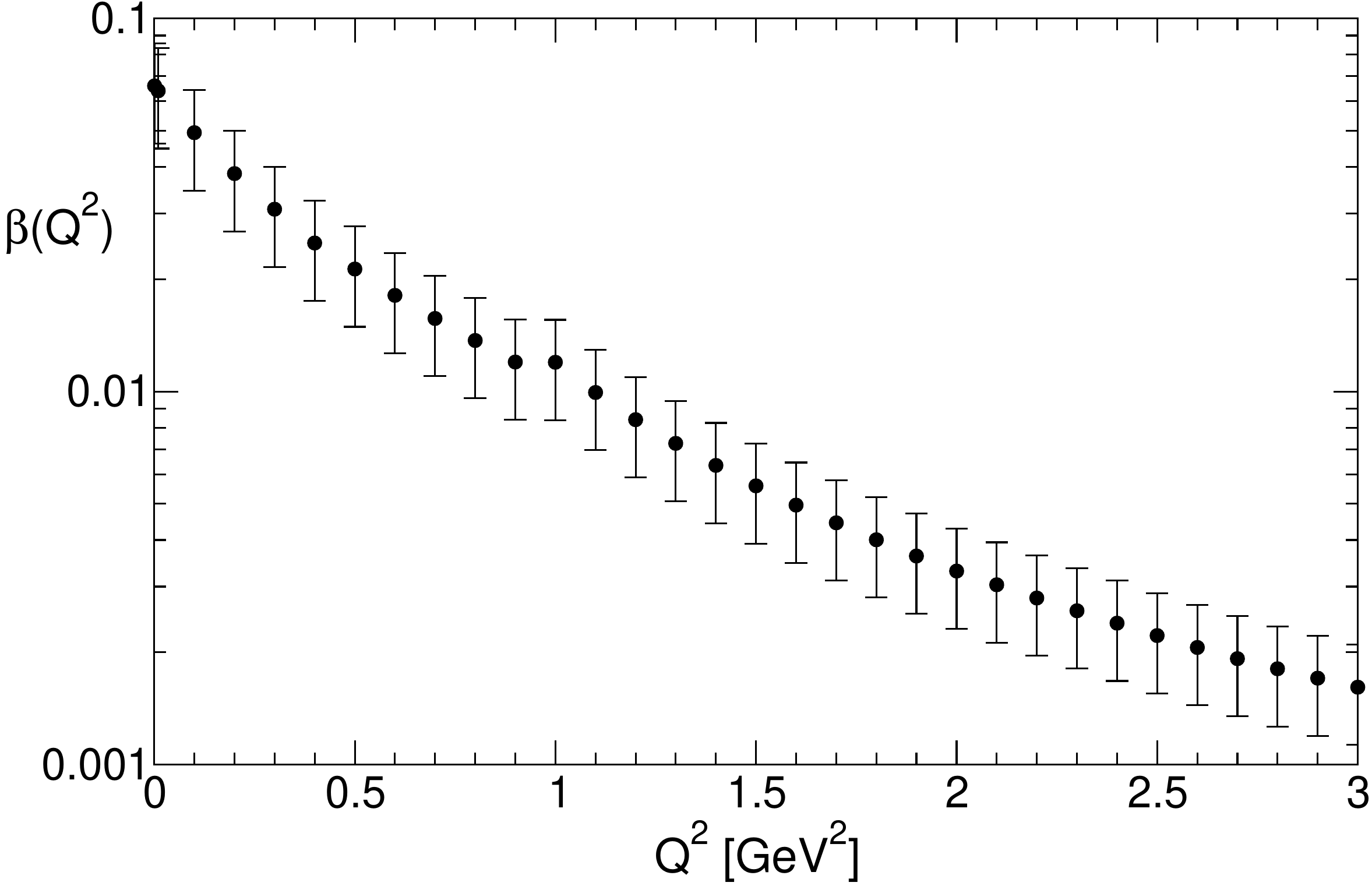} 
\caption{Residue  of the leading Reggeon contribution (\ref{eq:TR}). \label{fig:betaR} }
\end{figure}

For completeness, we show  in Fig. \ref{fig:betaR} the function $\beta(Q^2)$ for $Q^2$ in the range $(0, 3) \gev^2$, with an overall error of about 30\%. The figure shows that the two representations obtained from different sources for $Q^2<1 \gev^2$ and  $Q^2>1 \gev^2$ match remarkably well.   Note that the slightly different Regge expression given in Eq. (18) of \cite{Gasser:2020hzn} coincides at large $\nu$ with (\ref{eq:TR}) for $\beta(Q^2)=(2 m)^\alpha \beta_\alpha(Q^2)$, where $\beta_\alpha(Q^2)$ is the residue represented in Fig. 4 of \cite{Gasser:2020hzn}.
~\vspace{0.cm}

\section{Results}\label{sec:results}
\subsection{Lower bound $\delta_0^R$ at large energies}\label{sec:delta0R}
 As shown in Sect. \ref{sec:dual}, the lower bound  $\delta_0^R$ can be calculated by means of a numerical algorithm: it is given by the norm (\ref{eq:delta0})  of the Hankel matrix  ${\cal H}$, defined in  (\ref{eq:hank}) in terms of the Fourier coefficients $c_n$. The input consisting from the structure function and the Regge residue enters through the coefficients $c_n$, according to (\ref{eq:cn1}).  In the calculations, the matrix  ${\cal H}$ was truncated at a finite order, stability of the results being achieved with  about 30 Fourier coefficients $c_n$.

 Using this algorithm and the input described in the previous section, we calculated  $\delta_0^R$ for $Q^2$ in the range $(0, 3) \gev^2$.
 A free parameter is the energy $\nu_h$ above which we test the onset of the Regge asymptotics. In our analysis we increased this value from $W_h=3 \gev$ up to  $W_h=30 \gev$. The corresponding values of $\nu_h(Q^2)$ are obtained from (\ref{eq:W}). 
\begin{table}[htb]
\caption{Lower bound $\delta_0^R(Q^2)$ defined in (\ref{eq:delta0R})  as a function of $Q^2$ for several values of $W_h$.}\label{tab:1}
 \renewcommand{\tabcolsep}{0.55pc} 
\renewcommand{\arraystretch}{1.15} 
\begin{tabular}{|l|c|c|c| }\hline
$Q^2$ & $W_h=3 \gev$ & $W_h=10 \gev$ & $W_h=30 \gev$  \\\hline
0 & 0.1279 & 0.000594 &0.0000548 \\
0.01  & 0.1165  & 0.000582   &   0.0000538\\
0.1   & 0.0742&  0.000501  &   0.0000466  \\
0.2 & 0.0553	& 0.000433  &   0.0000406  \\
0.3 & 0.0342 &   0.000381  &    0.0000361  \\
0.4 & 0.0177   &  0.000344  & 0.0000326  \\
0.5 &0.00689	& 0.000310  &   0.0000298  \\
0.6 & 0.00063 &	0.000288 &   0.0000274  \\
0.7 & 0.00307	& 0.000266   &  0.0000255  \\
0.8 & 0.00458  &	0.000249    &  0.0000239  \\
0.9  & 0.00498 &    0,000234  &  0.0000224  \\
1 & 0.00476   &    0.000221  &  0.0000212  \\
1.5 &0.00311 &   0.000861     &  0.0000291  \\
2  &0.00110  &    0.000411   &  0.0000940  \\
2.5 &0.00047  &   0.000172  &   0.0000602  \\
3  & 0.00080  &   0.000053   &   0.0000095  \\
\hline \end{tabular}
 \vspace{0.1cm}
\end{table}

 For illustration, we give in Table \ref{tab:1} the values of the lower bound $\delta_0^R(Q^2)$ 
for three values of $W_h$, equal to 3, 10 and 30 GeV, respectively. Central values of the input have been used in the calculations. 

It is useful to compare the values given in Table \ref{tab:1} with the magnitude of the leading Reggeon. Recalling the definition of $\delta_0^R(Q^2)$, the comparison should be made with the quantity $|T_1^R|/\nu^\alpha$, which is independent on $\nu$ as follows from (\ref{eq:TR}).  Therefore, we show in Table \ref{tab:2} the values of the ratio 
\be\label{eq:r}
r(Q^2)=\frac{\delta_0^R(Q^2)}{|T_1^R|/\nu^\alpha}= \frac{\delta_0^R(Q^2)}{\pi\beta(Q^2)/\sin(\pi\alpha/2)},\ee 
for various $Q^2$ and increasing values of $W_h$. 
\begin{table}[htb]
\caption{Ratio $r(Q^2)$ defined in (\ref{eq:r}) as a function of $Q^2$ for several values of $W_h$.}\label{tab:2}
 \renewcommand{\tabcolsep}{0.55pc} 
\renewcommand{\arraystretch}{1.15} 
\begin{tabular}{|l|c| c|c| }\hline
$Q^2$& $W_h=3 \gev$  & $W_h=10 \gev$ & $W_h=30 \gev$  \\\hline
0 & 0.4681 & 0.00217 & 0.000201 \\
0.01 &0.4398  &   0.00221	&      0.000201\\
0.1  &0.3632   &  0.00245	&      0.000228 \\
0.2 & 0.3479	&  0.00273	&	  0.000256 \\
0.3 & 0.2683 &    0.00301	&	      0.000280 \\
0.4  &0.1693  &  0.00328	&      0.000311 \\
0.5 & 0.0785 &	0.00355		& 	0.000339 \\
0.6 & 0.0085	& 0.00384	&	0.000366 \\
0.7 & 0.0474&	0.00412	 &		0.000394 \\
0.8 & 0.0809 &	0.00440		&	0.000422 \\
0.9 & 0.0998  &    0.00468	&      0.000449 \\
1 & 0.1069    &     0.00496	&         0.000477 \\
1.5  & 0.1347 &   0.0375	&	      0.00126 \\
2 &0.0835     &   0.0311	&	        0.00710 \\
2.5 & 0.0531   &   0.0190	&	       0.00669 \\
3   & 0.1180   &   0.0079	&         0.00141 \\
\hline \end{tabular}
 \vspace{0.1cm}
\end{table}

From Tables \ref{tab:1} and \ref{tab:2} it can be seen that for each $Q^2$ the values of  $\delta_0^R(Q^2)$ and $r(Q^2)$ become smaller when $W_h$ is increased: while for $W_h=3 \gev$ the lower bound is rather large and represents for some $Q^2$ a substantial fraction of the dominant Reggeon, these values decrease in an impressive way with increasing $W_h$.

One might ask how these results can be interpreted.  The existence of a lower bound shows that the difference between the exact amplitude and the contribution of the dominant Reggeon must exceed a certain value in order to ensure consistency with the low-energy structure function. A large value of the lower bound, of the order of magnitude of the dominant Reggeon contribution for instance, would question the validity of the Regge asymptotics or would suggest that large corrections to the Regge asymptotics are required by the low-energy data and analyticity. 

Our results  show however that this is not the case: the lower bound on the difference between the true amplitude and the leading Reggeon contribution  decreases when the energy is increased and becomes a tiny fraction of the Reggeon contribution.

 From Tables \ref{tab:1} and \ref{tab:2}, one can see that the decrease is impressive   for $Q^2\leq 1 \gev^2$, when the parametrization  of the structure function from \cite{Alwall:2004wk} is used as input for $W>3 \gev$. Moreover, for $Q^2\leq 1 \gev^2$, the quantities $\delta_0^R(Q^2)$ and $r(Q^2)$ continue to rapidly decrease when $W_h$ is increased above 30 GeV. 

For $Q^2>1 \gev^2$, it turns out that the decrease of the quantities  $\delta_0^R(Q^2)$ and $r(Q^2)$ is slowed down when $W_h$ is further increased above 30 GeV. This may be due to the less  precise input, in particular to the fact that  the parametrization \cite{Alekhin:2013nda} that we used for $Q^2>1 \gev^2$ and $W> 3 \gev$ contains a small number of points in the region of interest for our calculations. Further investigations of this region using the available parametrizations of the structure functions are necessary.

As emphasized in Sect. \ref{sec:dual}, the lower bound (\ref{eq:delta0R}) is saturated, i.e. there exists an analytic function, which satisfies the constraints, and for which the difference (\ref{eq:diffnu}) is equal to $\delta_0^R$.   The small values of $\delta_0^R(Q^2)$ given in Table \ref{tab:1} show that, if the lower bound would be saturated by the exact amplitude $T_1^\inel$, there would be little room for subasymptotic corrections, in particular for the contribution of a fixed Regge pole, in the asymptotic behaviour of this amplitude. This would support the Reggeon dominance hypothesis (\ref{eq:RD}), adopted in \cite{Gasser:1974wd, Gasser:2015dwa, Gasser:2020hzn, Gasser:2020mzy}.

 Of course, the  difference between the true amplitude and the Reggeon contribution is not in principle limited from above. So, a strong statement about the validity of the Reggeon dominance hypothesis cannot be made.  In the same time,  the approximate saturation of the lower bound is not excluded.  Therefore, having in view the low values of  $\delta_0^R$ given in Table \ref{tab:1}, we can  make the conservative statement that the data on the structure function are consistent with the Reggeon dominance hypothesis.  

\subsection{Determination of the subtraction function}\label{sec:S1}
 We consider now the lower bound $\widehat\delta_0^R(S_1^\inel)$ defined in (\ref{eq:delta0Rhat}), which measures the difference between the true amplitude and the dominant Reggeon when the value of the amplitude at $\nu=0$ is specified as in (\ref{eq:S1}). 

 As shown in Sect. \ref{sec:dual}, the lower bound  $\widehat\delta_0^R$ is given by the norm (\ref{eq:delta0hat})  of the Hankel matrix $\widehat {\cal H}$, defined in (\ref{eq:hankhat})in terms of the Fourier coefficients $\widehat c_n$, which depend on the input consisting from  the structure function, the Regge residue and the parameter $S_1^\inel(Q^2)$, according to (\ref{eq:cnhat}). 

We recall that $S_1^\inel(Q^2)$ is usually denoted as subtraction function, as it enters the subtracted dispersion relation (\ref{eq:DRinel}). 
As  we discussed in Section \ref{sec:introduction}, assuming Reggeon dominance (\ref{eq:RD}), one can derive a sum rule for the subtraction function in terms of the structure function along the whole unitarity cut in the $\nu^2$ plane. This sum rule, written down for the first time in \cite{Elitzur:1970yy}, was derived independently in  \cite{Gasser:1974wd}  and was applied more recently in \cite{Gasser:2015dwa, Gasser:2020hzn, Gasser:2020mzy} for the determination of the electromagnetic neutron-proton mass difference and nucleon polarizabilities.

\begin{figure}[htb]
\centering \vspace{0.2cm}
\includegraphics[width=\linewidth, width=4cm]{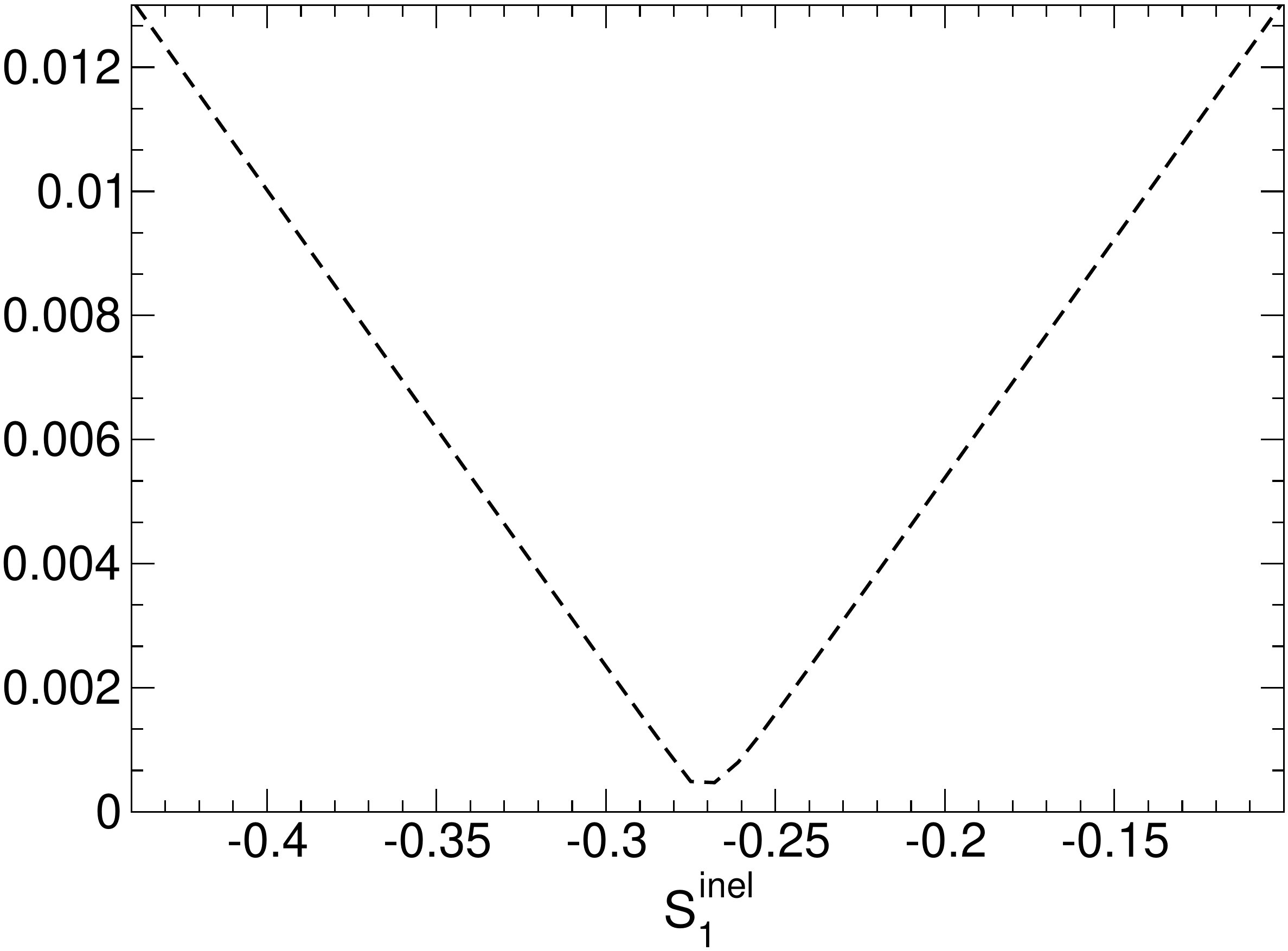} \hspace{0.2cm} \includegraphics[width=\linewidth, width=4cm]{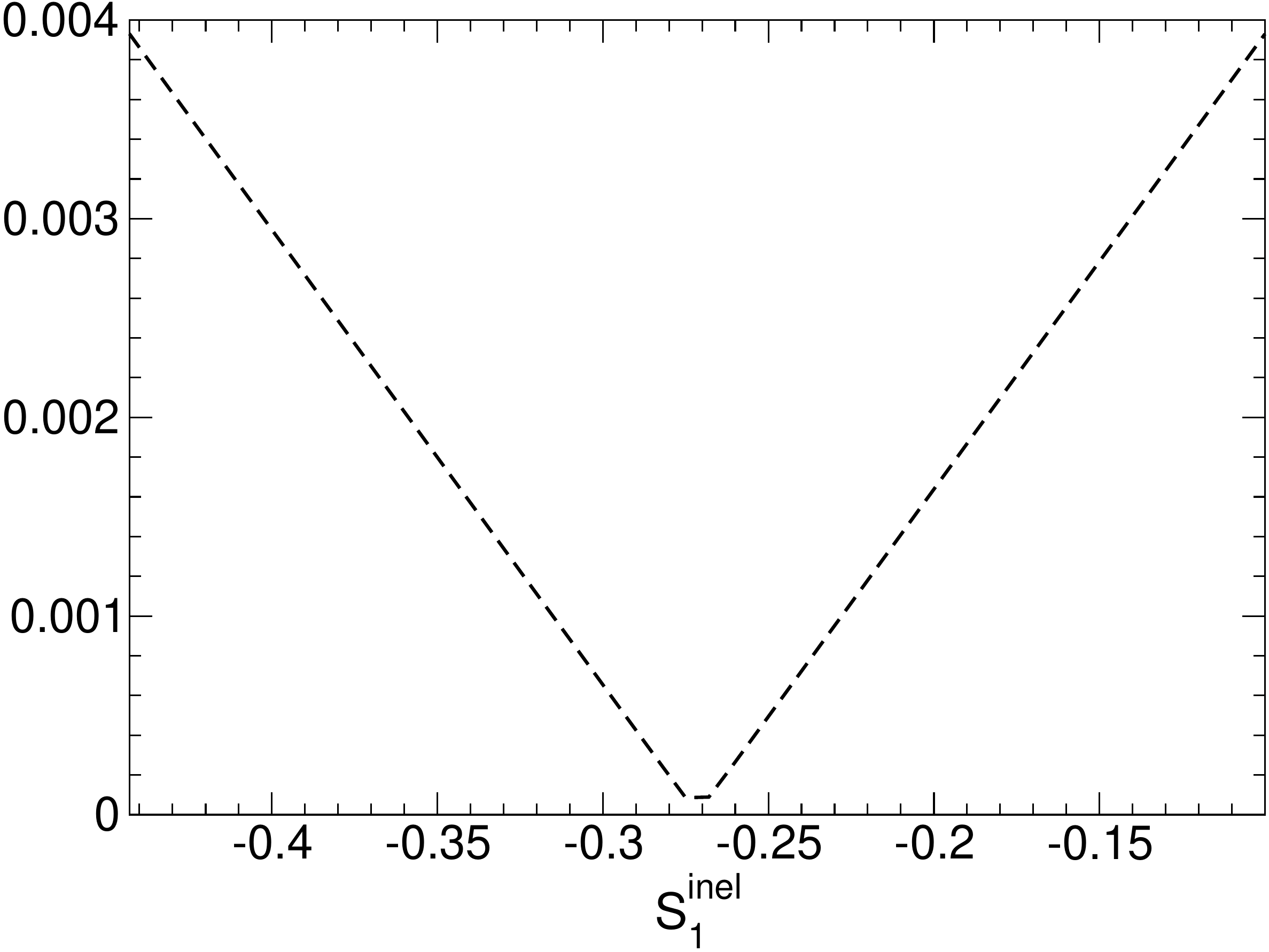} 
\caption{Lower bound $\widehat \delta_0^R(S_1^\inel)$ as a function of $S_1^\inel$  for $Q^2=0.3 \gev^2$. Left: $W_h=10 \gev$; right: $W_h=30 \gev$. \label{fig:deltaS1}}
\end{figure}

One might ask whether the subtraction function can be predicted  by imposing Regge asymptotics in a way less stringent than the hypothesis (\ref{eq:RD}). An answer to this question was given in \cite{Caprini:2016wvy},  where bounds on the subtraction function have been derived by assuming that  for  $W>3 \gev$, along the cut, the modulus of the amplitude is described by the Regge model, including subleading contributions. This approach led to results for $S_1^\inel$  consistent, but weaker than those reported in \cite{Gasser:2015dwa}. 

In the present paper, we argue that the lower bound $\widehat\delta_0^R(S_1^\inel)$ offers an alternative tool for estimating the subtraction function, which is similar in spirit, but not identical with the Reggeon dominance hypothesis (\ref{eq:RD}). 
Namely, we make the conjecture that, if Regge asymptotics is valid in virtual Compton scattering,  the value of $S_1^\inel$ which achieves the minimum of the lower bound $\widehat\delta_0^R(S_1^\inel)$ can be taken as the best estimate of the subtraction function. 

The reasoning behind this conjecture is simple: if Regge asymptotics is assumed to be valid, the difference (\ref{eq:diffnu}) for high enough $\nu_h$ should be as small as  possible  for the exact amplitude and the exact value $S_1^\inel$ of this amplitude at $\nu=0$. The same will be true for the minimal difference (\ref{eq:delta0Rhat}), which by definition is smaller than the exact difference. Moreover, if the lower bound is small,  the exact difference (\ref{eq:diffnu}) can also be small, which would not be true in the opposite case of a large lower bound. It is then natural to determine the unknown $S_1^\inel$ by looking for the minimum of the lower bound  $\widehat\delta_0^R(S_1^\inel)$ on the difference (\ref{eq:diffnu}).

The conjecture is also supported by the behaviour of the lower bound $\widehat\delta_0^R(S_1^\inel)$ as a function of $S_1^\inel$: it turns out that for all $Q^2$, this quantity  exhibits a sharp minimum for a certain value of  $S_1^\inel$. Moreover, the minimum is stable when the energy $W_h$ is increased. These properties  are illustrated in Fig. \ref{fig:deltaS1}, where we show the lower bound $\widehat\delta_0^R(S_1^\inel)$  for $Q^2=0.3 \gev^2$ as a function of   $S_1^\inel$ for $W_h=10 \gev$ and $W_h=30 \gev$. It is remarkable that, although the values of the lower bound are different (being smaller for higher $W_h$, in agreement with the results of the previous subsection), the minimum occurs at the same value of  $S_1^\inel$. This stability holds very precisely for all $Q^2\leq 1 \gev^2$ and, within some uncertainties, also for  $Q^2 > 1 \gev^2$. Therefore, the minimum of  $\widehat\delta_0^R(S_1^\inel)$ offers a well-defined and unambiguous definition of the subtraction function.

\begin{figure}[htb]
\centering \vspace{0.2cm}
\includegraphics[width=\linewidth, width=7cm]{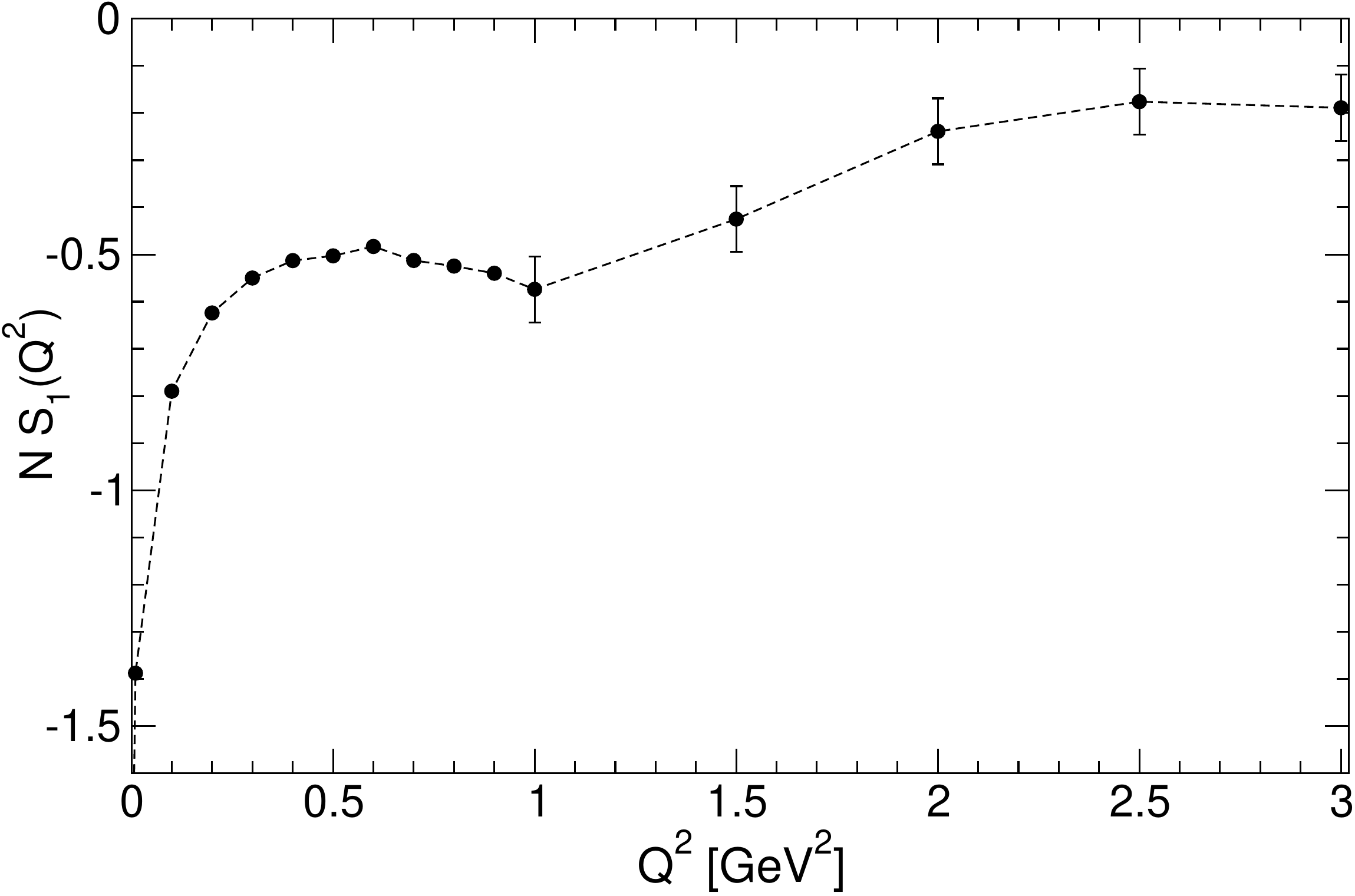} 
\caption{Product $N S_1^\inel(Q^2)$ in units of $\gev^{-2}$, for $0\leq Q^2\leq 3$, where $N=(1+Q^2/M_d^2)^2$, with $M_d^2=0.71 \gev^2$. \label{fig:S1}}
\end{figure}

In Fig. \ref{fig:S1} we present the values of $S_1^\inel(Q^2)$ determined by this procedure for $0\leq Q^2\leq 3 \gev^2$. For convenience, as in \cite{Gasser:2015dwa, Caprini:2016wvy}, we show the product of $S_1^\inel$ with the factor $N=(1+Q^2/M_d^2)^2$, with $M_d^2=0.71 \gev^2$. The behaviour around the point $Q^2=1 \gev^2$ is explained by the different input above $W=3 \gev$  used below and above this point: for $Q^2\leq 1 \gev^2$,
 the data are taken from the parametrization given in \cite{Alwall:2004wk}, while for $Q^2>1 \gev^2$, we use the data from \cite{Alekhin:2013nda}.  In the latter case, the stability of the minimum of the lower bound when $W_h$ is increased occurs within some limits. For illustration, the values  shown in Fig. \ref{fig:S1} are the average of the determinations at $W_h=10 \gev$ and $W_h=30 \gev$, with an uncertainty equal to their difference. 

The values presented in Fig. \ref{fig:S1} are obtained with the central values on the experimental data on the structure function $V_1$ and the Regge model $T_1^R$. Since the purpose of this paper was mainly to introduce the new formalism, we briefly discuss below  the effect of  the uncertainties only in the particular case $Q^2=0$, deferring a full analysis to a future work.

The subtraction function at $Q^2=0$  satisfies the low-energy theorem  \cite{Gasser:2015dwa}
\beq\label{eq:S1in0}
  S_1^\inel(0)= - \frac{\kappa^2}{4\mN^2}-\frac{\mN}{\alphaem}\,\beta_M,
\eeq
where  $\alphaem$  is the fine structure constant, $\kappa$ is the anomalous magnetic moment of the particle and $\beta_M$ its  magnetic polarizability. We recall that we consider the difference of the proton and neutron amplitudes, so the theorem involves the difference of these parameters for the proton and neutron. 

From Fig. \ref{fig:S1}, one can read the central value  $S_1^\inel(0)=-0.745 \gev^{-2}$ obtained from our prescription. By varying the residue of the Reggeon contribution (\ref{eq:TR}) by $\pm 30\%$, the position of the  minimum of the lower bound is shifted  by $\pm 1.295$. A conservative error of $30\%$ on the structure function leads to another variation of the central value by $\pm 1.065$. Adding the errors in quadrature, we obtain
\be\label{eq:S1res}
  S_1^\inel(0)= (-0.745 \pm 1.677),
\ee  
in $\gev$ units. For comparison, we quote the previous predictions based on Regge asymptotics, 
 $S_1^\inel(0)=0.3\pm 1.2$ reported in \cite{Gasser:2015dwa} and  $-1.59 \leq S_1^\inel(0) \leq 1.22$ derived in \cite{Caprini:2016wvy}, and the experimental estimate $S_{1, \text{exp}}^\inel(0)=1 \pm  2.7$, given in \cite{Gasser:2015dwa}. Although the central value in  (\ref{eq:S1res} is somewhat lower, our result is  consistent with the previous predictions within the errors.

One can check that the results for $S_1^\inel(Q^2)$  shown in Fig.  \ref{fig:S1} are consistent also with the upper and lower bounds on the subtraction function in the range $0<Q^2\leq 1 \gev^2$, shown in Fig. 3 of \cite{Caprini:2016wvy}. Moreover, our results agree with the predictions of Reggeon dominance hypothesis, represented by the uncertainty band shown in Fig. 7 of  \cite{Gasser:2015dwa} for $0.5\leq Q^2\leq 1 \gev^2$.\footnote{In  \cite{Gasser:2015dwa} the uncertainty
band was not represented for  $Q^2 < 0.5 \gev^2$, because the results are there sensitive to the unphysical fluctuations in the parametrizations of the data.}.  We shall briefly discuss  the significance of this result in Sect. \ref{sec:conclusions}.

We note finally that for  the values  shown in Fig.  \ref{fig:S1} for $Q^2>1 \gev^2$, the comparison with the results given in   \cite{Gasser:2015dwa, Caprini:2016wvy} for $Q^2>1 \gev^2$ cannot be made, since these works used a different input for $W>3 \gev$. Further investigations of this region are actually necessary, using updated experimental data on the structure functions.

\section{Summary and conclusions}\label{sec:conclusions}
The purpose of this work was to investigate the validity of Regge asymptotics for the amplitude of virtual Compton scattering on the nucleon. Specifically, the aim was to check whether the data on the nucleon structure function, used as input up to a certain energy $\nu_h$, are consistent with a behaviour given by Reggeon exchanges for  $\nu\ge \nu_h$. 

The work was motivated by the present applications of dispersion theory in  Compton scattering. Since one of the invariant amplitudes,  $T_1(\nu, Q^2)$,  increases at high $\nu$, the dispersion relation at fixed $Q^2$ requires a subtraction. Therefore, the amplitude cannot be recovered completely from the knowledge of the structure function known phenomenologically on the unitarity cut, but depends also on the  subtraction function $S_1(Q^2)$, i.e. the value of the amplitude at $\nu=0$. 

In general, if the asymptotic behaviour of the amplitude is not known in detail,  the subtraction constants cannot be calculated and are viewed as independent ingredients in the dispersion relation. For Compton scattering, this view was adopted in \cite{WalkerLoud:2012bg, WalkerLoud:2012en, Erben:2014hza, Thomas:2014dxa, Walker-Loud2018}.

However, if one assumes that  Regge asymptotics is valid,  the subtraction function can be determined in terms of the structure function  by exploiting in a suitable way the known  behaviour. Using the Reggeon dominance hypothesis  (\ref{eq:RD}),  a sum rule for the subtraction function in terms of the structure function was derived independenly a long time ago in  \cite{Elitzur:1970yy} and \cite{Gasser:1974wd}, and was evaluated with modern input in the recent analyses  \cite{Gasser:2015dwa, Gasser:2020hzn, Gasser:2020mzy}.

The validity of Regge behaviour at high energies is therefore important for improving the precision  of dispersion theory for the Compton amplitude. 
In the present paper, we proposed a test of  Regge asymptotics  using methods of functional analysis. For illustration,  we considered the  difference of the inelastic amplitudes $T_1^\inel(\nu, Q^2)$ relevant for proton and neutron. 

The central role in our analysis is played by the quantities  $\delta^R_0$ and  $\widehat \delta^R_0$, defined as the solutions of the functional minimization problems (\ref{eq:delta0R}) and (\ref{eq:delta0Rhat}).  They represent lower bounds on the maximum difference between the exact amplitude $T_1^\inel$ and the Reggeon contribution  $T_1^R$ above a sufficiently high energy $\nu_h$,  imposed by analyticity and the structure function $V_1$  known  for $\nu< \nu_h$.
As shown in appendix \ref{sec:solution}, the lower bounds can be calculated by a numerical algorithm. 

The  results given in Tables \ref{tab:1} and \ref{tab:2} show that, for each $Q^2$ in the range $(0, 3) \gev^2$,  $\delta^R_0$ exhibits an impressive decrease when the energy $W_h$ is increased, and becomes only a tiny fraction of the leading Reggeon contribution.

As discussed in Sect. \ref{sec:results},  large values of the lower bound at high energies would put doubts on the validity of the Regge asymptotics. However,  the small values of the lower bound given in Table \ref{tab:1}  show that this is not the case. If  the true amplitude $T_1^\inel$ would saturate the lower bound, there would be little room for subasymptotic corrections in the asymptotic behaviour of this amplitude.
 We note in particular that a fixed Regge pole at $J=0$ would contribute to the difference (\ref{eq:diffnu}) as a term $C/\nu_h^\alpha$, with real $C$. If the lower bound would be saturated by the exact amplitude $T_1^\inel$, neglecting other subleading contributions to this amplitude, one would have $\delta_0^R\sim C/\nu_h^\alpha$, allowing the extraction of the residue $C$. But, since the residue is not dimensionless, its values do not have a direct meaning. The magnitude of the fixed-pole contribution can be controlled  by comparing it with the corresponding contribution of the dominant Reggeon. This amounts to consider the ratio of the quantities $C/\nu_h^\alpha$ and $|T^R_1|/\nu_h^\alpha$.  Assuming that the lower bound is saturated, i.e. $C/\nu_h^\alpha\sim \delta_0^R$, it follows that  an appropriate dimensionless parameter that controls the magnitude of the fixed-pole contribution is the ratio $r$ defined in (\ref{eq:r}). The values of this quantity, presented in Table \ref{tab:2}, show that the fixed pole brings a negligible contribution compared to the leading Regge term in the range of $Q^2$ considered.

 Of course, the existence of a lower bound does not limit from above the  difference between the exact amplitude and the Reggeon contribution. So, a strong statment about the contribution of a fixed pole cannot be made. On the other hand, the small value of the lower bound mean that a small value of the exact difference is not forbidden. Therefore, the results given in Tables \ref{tab:1} and  \ref{tab:2} allow us to make the conservative statement that the data on the structure function are consistent with the standard Regge asymptotics.  

Finally, in Section \ref{sec:S1} we proposed a prescription for obtaining the subtraction function $S_1^\inel(Q^2)$ in a frame consistent with Regge asymptotics. Specifically, our conjecture is to take the value which minimizes the lower bound $\widehat \delta_0^R$ defined in (\ref{eq:delta0Rhat}). The behaviour of $\widehat \delta_0^R$  as a function of $S_1^\inel(Q^2)$ and the stability of the minimum when $W_h$ is increased,  shown in Fig. \ref{fig:deltaS1}, support this conjecture. 
Remarkably,  for $0.5 \leq Q^2\leq 1 \gev^2$ the results presented in Fig. \ref{fig:S1}  are very close to the central values of the uncertainty band  obtained in \cite{Gasser:2015dwa} from the Reggeon dominance hypothesis  (\ref{eq:RD}).

 The significance  of this result deserves attention, having in view that in our formalism Regge asymptotics is implemented  a way similar, but not identical to  the Reggeon dominance hypothesis. Indeed, both approaches use as input the imaginary part of the amplitude along the cut up to a certain $\nu_h$, and the behaviour of the full amplitude above $\nu_h$. However, the Reggeon dominance hypothesis refers to the asymptotic limit $\nu_h\to \infty$, while in the present formalism $\nu_h$ is kept finite. Moreover, the Reggeon dominance assumes that the difference between the true amplitude and the Reggeon contribution is strictly zero at $\nu\to\infty$, while in the present approach one finds a lower bound on this difference for $\nu\ge \nu_h$.   One might think
that the Reggeon dominance hypothesis is somehow obtained as a limit of the present formalism 
for large $\nu_h$. However, the limit $\nu_h\to\infty$ cannot be defined in a strict mathematical sense in the present formalism, which assumes explicitely that  $\nu_h$ is finite.   

 As mentioned above,  the position of the minimum of  $\widehat \delta_0^R$ as a function of  $S_1^\inel(Q^2)$   is stable, while the corresponding value of $\widehat\delta_0^R$ decreases when $W_h$ is increased. This trend continuous in a spectacular way when $W_h$ is further increased up to high values of about 10000 GeV:  the minimum occurs at the same place, whereas the corresponding values of  $\widehat\delta_0^R$ and $|T_1^\inel-T_1^R|$ become smaller and smaller. As mentioned above, in our formulation of the problem,  the limit $\nu_h\to\infty$  cannot be defined in a formal way. However, the numerical trend\footnote{Numerical instabilities are expected to occur at very large $\nu_h$, since the exact limit $\nu_h\to\infty$ does not exist.} suggests that the difference $|T_1^\inel-T_1^R|$ approaches  extremely small values, consistent with the Reggeon dominance hypothesis (\ref{eq:RD}). This explains why the sum rule derived from Reggeon dominance in \cite{Gasser:2015dwa, Gasser:2020hzn} and the prescription proposed in this work, although mathematically quite different, lead to close predictions for the subtraction function. 

We emphasize that the present work has an exploratory character: the aim was to propose a new approach for testing Regge asymptotics and its implications.  For illustration we considered the amplitude $T_1^\inel$ and used only central values of the phenomenological input.  The investigation of the amplitude $\bar T(\nu, Q^2)$, used recently in \cite{Gasser:2020hzn, Gasser:2020mzy}  for the evaluation of Cottingham formula,  and a detailed analysis of the uncertainties, will be considered in a future work. 

\vspace{-0.1cm}

\subsection*{Acknowledgments}
I thank H. Leutwyler for very useful discussions. This work was supported by the Romanian Ministry of Education and Research,  Contract PN 19060101/2019-2022.

\appendix
\section{Solution of the extremal problems}\label{sec:solution}
We treat in detail the extremal problem  (\ref{eq:delta0R}), the solution of (\ref{eq:delta0Rhat}) being obtained from it by straightforward modifications.  In Sect. \ref{sec:canon} we express the problem in a canonical form and in Sect. \ref{sec:dual} we solve it by applying a duality theorem in optimization theory \cite{Duren}. 
We mention that these techniques have been applied already in other physical problems - for a review see \cite{Caprini:2019osi}.

\subsection{Canonical formulation}\label{sec:canon}

In (\ref{eq:delta0R}), the relevant function appears to be the ratio $T_1^\inel/\nu^\alpha$. But the simple idea to define a new function by dividing the amplitude $T_1^\inel$ by $\nu^\alpha$ does not work, since this introduces an unwanted singularity at $\nu=0$. Since (\ref{eq:diffnu}) involves only the modulus  $|\nu^\alpha|$ for $\nu^2\ge \nu^2_h(Q^2)$,  we must define a function analytic in the $\nu^2$ plane cut for  
 $\nu^2 \ge \nu^2_h(Q^2)$, such that its modulus on this cut is equal to $\nu^\alpha$.
Moreover, the function should not have  zeros in the complex plane, since they would affect the optimality of the bound \cite{Duren}. This means that we must construct a so-called ''outer function'', which has the desired modulus on the cut, and does not have zeros in the complex  plane. 

\begin{figure}[htb]
\centering \vspace{0.2cm}
\includegraphics[width=\linewidth, width=3.5cm]{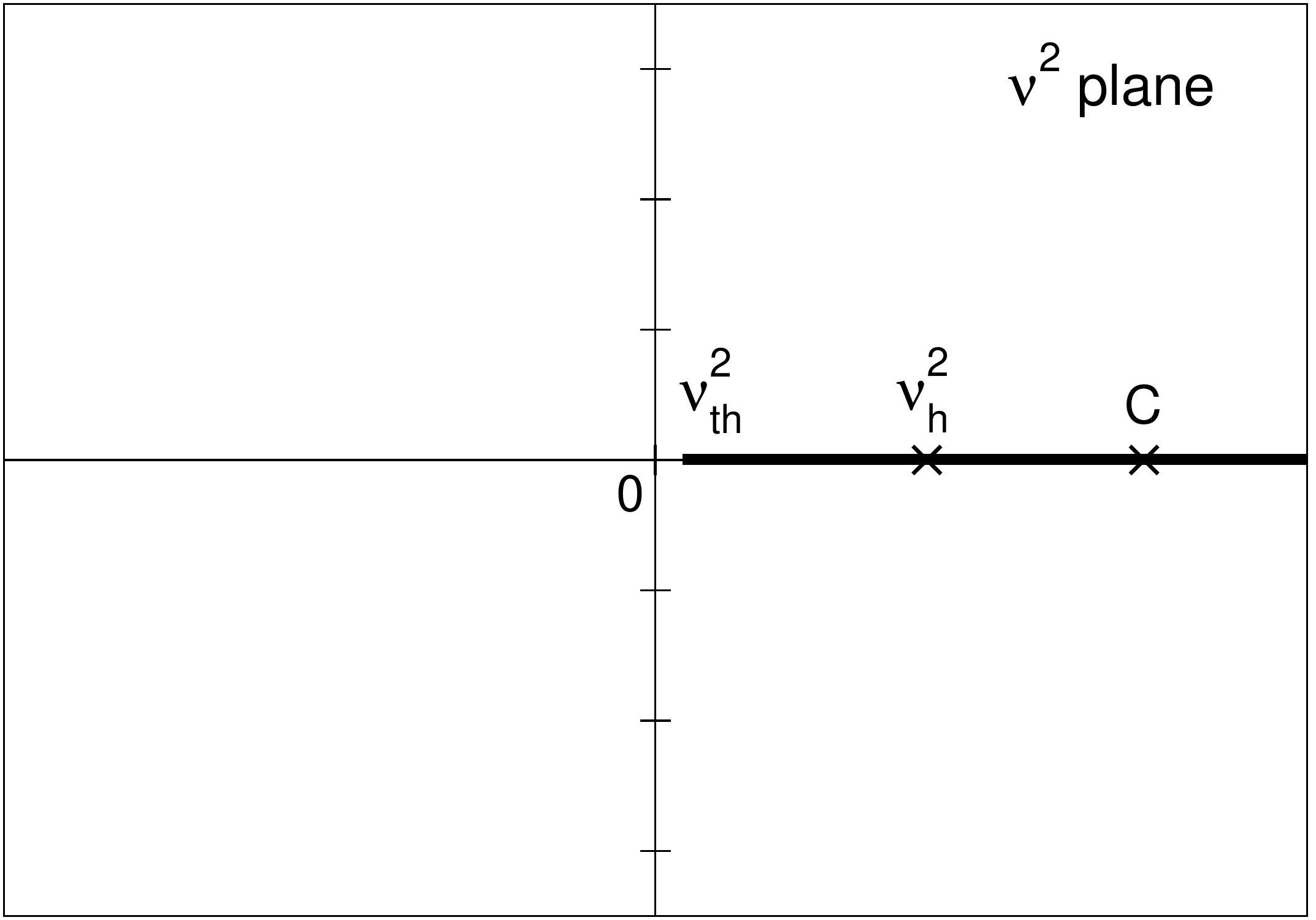} \hspace{1cm} \includegraphics[width=\linewidth, width=3.5cm]{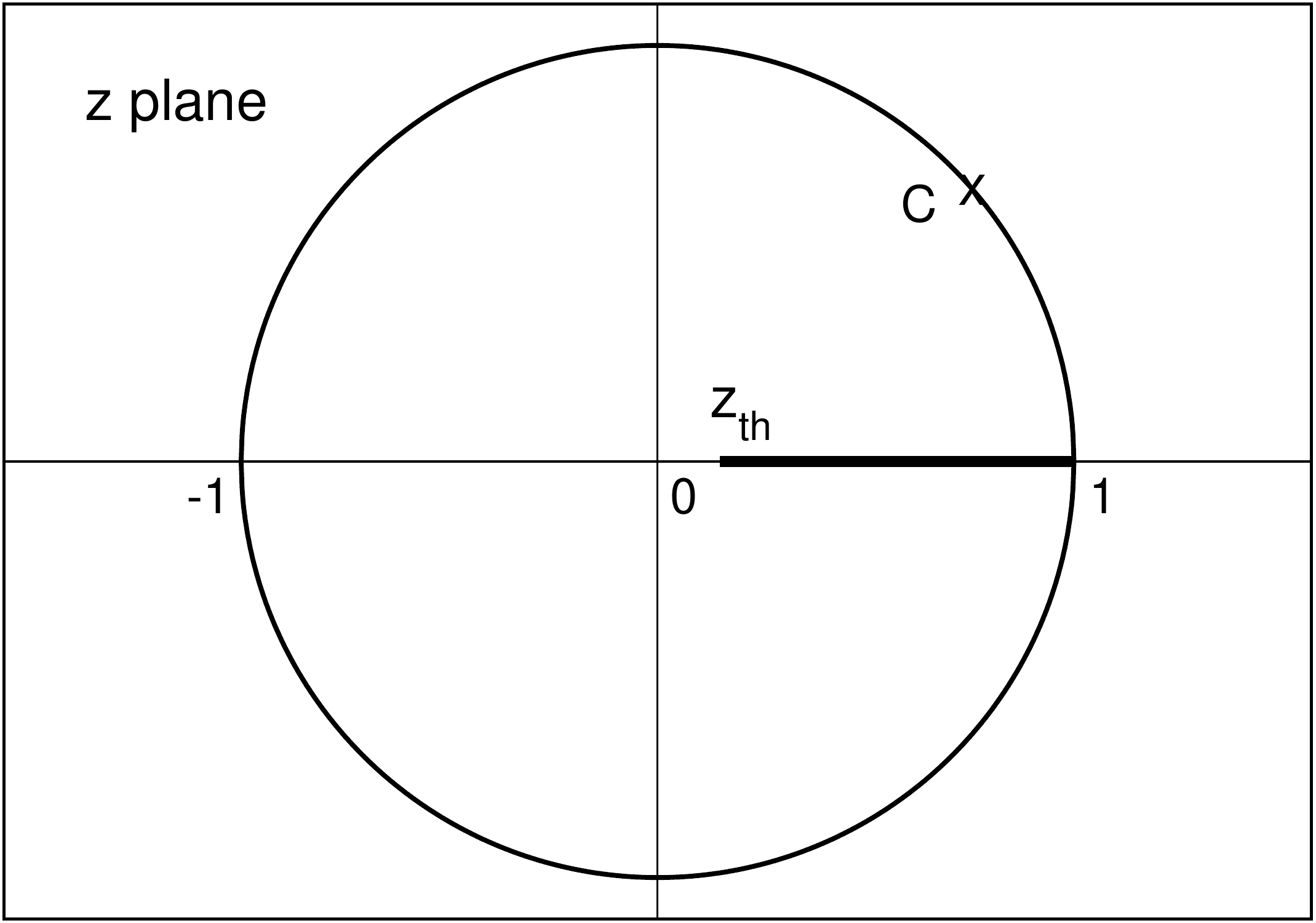} 
\caption{Conformal mapping (\ref{eq:z}). Left: $\nu^2$ plane. Right: $z$ plane. \label{fig:z}}
\end{figure}

As shown in \cite{Duren}, an outer function is defined in general by an integral representation. However, in our case we can obtain the required function in a closed form by a simpler technique  \cite{Caprini:2019osi}. We make first the  change of variable 
 \beq\label{eq:z}
z\equiv \tilde z(\nu^2)=\frac{1-\sqrt{1-\nu^2/\nu_h^2}}{1+\sqrt{1-\nu^2/\nu_h^2}},
\eeq 
which performs the conformal mapping of the $\nu^2$ plane onto the interior of the unit disk $|z|<1$,  such that the upper (lower) edge of the cut along  $\nu^2\ge \nu_h^2$  becomes the upper (lower) semicircle of the circle $|z|=1$.  We note also that  $\tilde z(0)=0$, and the cut for $ \nu_{\thr}^2 \leq \nu^2 \leq \nu_h^2 $ becomes a cut along the segment $z_{th}\le z\le 1$ in the $z$-plane, where 
\beq\label{eq:zth}
z_{\thr}=\tilde z(\nu_{\thr}^2).
\eeq
The mapping is ilustrated schematically in Fig. \ref{fig:z}, which shows that the segment $(\nu^2_{th}, \nu_h^2)$ of the $\nu^2$ plane is mapped onto the segment $(z_{th}, 1)$ of the $z$ plane, and a point $C$ on the upper rim of the cut above $\nu^2_h$ is mapped onto a point $C$ on the circle $|z|=1$. 

It is useful to consider also the inverse of $\widetilde z(\nu^2)$, which has the form:
\beq\label{eq:tildenu}
 \nu^2\equiv\widetilde{\nu^2}(z)=\frac{4 z}{(1+z)^2}\,\nu_h^2.
\eeq
This function has a zero at $z=0$, which is the image of the origin $\nu^2=0$ of the original plane. By removing this zero, we define the function
\beq\label{eq:o}
o(z)=\frac{4}{(1+z)^2}\,\nu_h^2,
\eeq
which is analytic and without zeros in $|z|<1$. From (\ref{eq:tildenu}) and (\ref{eq:o}) if follows  that the ratio
\beq\label{wq:oz}
\frac{\nu^2}{o(z)}=z
\eeq
has modulus equal to 1 for $|z|=1$, i.e. for $\nu^2\ge \nu_h^2$.  Therefore,   $|o|=\nu^2$ for $\nu^2\ge \nu_h^2$.
Then, it is easy to see that the function
\be\label{eq:O}
O(z)\equiv [o(z)]^{\alpha/2}=\frac{2^\alpha}{(1+z)^\alpha}\,\nu_h^\alpha,
\ee
is the outer function which we looked for: indeed, it is analytic and without zeros in $|z|<1$, i.e. in the $\nu^2$ plane cut for $\nu^2\ge \nu_h^2$,  and  $|O|=\nu^\alpha$ along this cut.

We define now the functions 
\beq\label{eq:F}
F(z)\equiv \frac{T_1^\inel([\widetilde{\nu^2}(z)]^{1/2}, Q^2)}{O(z)}
\eeq
and
\beq\label{eq:FR}
F^R(z)\equiv \frac{T_1^R([\widetilde{\nu^2}(z)]^{1/2}, Q^2)}{O(z)}.
\eeq
Then, the supremum defined in (\ref{eq:diffnu}) can be written in an equivalent way as
\beq\label{eq:diffR}
\sup_{\theta\in (0,2\pi)} \vert F(e^{i\theta})- F^R(e^{i\theta}) \vert\equiv \Vert F-F^R\Vert_{L^\infty},
\eeq
where both functions $F(z)$ and $F^R(z)$ are bounded on the circle $|z|=1$. In (\ref{eq:diffR}) we introduced the $L^\infty$ norm \cite{Duren}, defined as the supremum of the modulus of the function
along the boundary $\vert z\vert=1$ of the unit disk. 

The analytic properties of the amplitude $T_1^\inel$ and of the function $o(z)$ imply that $F(z)$ is real analytic and bounded in the unit disk  $|z|< 1$, cut along the real segment $(z_{\thr}, 1)$.  The discontinuity across this cut is proportional to the imaginary part
\beq\label{eq:ImF}
\text{Im} F(y+i\epsilon)=\sigma(y), \quad z_{\thr}\leq y \leq 1, \eeq
where
\beq\label{eq:sigma}
  \sigma(y)\equiv \pi V_1(\widetilde{\nu^2}(y), Q^2)\,\frac{(1+y)^\alpha}{(2\nu_n)^\alpha}.
\eeq
It follows that we can write
\beq\label{eq:g}
F(z)= \frac{1}{\pi}\int_{z_{\thr}}^{1+\eta}\frac{\sigma(y)}{y-z}dy +  g(z),
\eeq 
where $\eta>0$ is a small positive number and $\sigma(y)$ for $y\ge 1$ is an arbitrary function, such that  $\sigma(y)$ is  continuous at $y=1$. As will be seen below, the results will not depend on $\eta$ and this freedom. From (\ref{eq:g}) it follows also that  the function $g(z)$  is free of constraints, and is therefore an arbitrary  holomorphic function bounded in $|z|<1$.

We define now the function $h(\zeta)$ for $\zeta=\exp(i\theta)$, i.e. on the boundary of the unit disk, as
\beq\label{eq:h}
h(\zeta)=  F^R(\zeta)-\frac{1}{\pi}\int_{z_{\thr}}^{1+\eta}\frac{\sigma(y)}{y-\zeta}dy.
\eeq
Then the lower bound (\ref{eq:delta0R}) can be written as
\begin{equation}\label{eq:inf1}
\delta_0^R=\min_{g\in H^\infty} \Vert g-h\Vert_{L^\infty},
\end{equation}
where  the minimization is with respect to all the functions $g(z)$
analytic  in the disk $\vert z \vert<1$ and bounded on its boundary (this class of functions is  denoted as $H^\infty$ \cite{Duren}).

Before presenting in the next subsection the solution of this problem, let us briefly indicate the modifications in the case of the more general problem (\ref{eq:delta0Rhat}).
Using (\ref{eq:S1}), we obtain from (\ref{eq:F})
\beq\label{eq:F0}
F(0)=\frac{S_1^\inel(Q^2)}{O(0)} =\frac{S_1^\inel(Q^2)}{(2\nu_h)^\alpha}.
\eeq
 We write now the most general representation of the function $F(z)$, which implements the constraint at $z=0$, as  
\beq\label{eq:g1}
F(z)= F(0)+\frac{z}{\pi}\int_{z_{\thr}}^{1+\eta}\frac{\sigma(y)}{y(y-z)}dy + z g(z),
\eeq
where $g(z)$ is an arbitrary bounded analytic function in $|z|<1$. Using this representation in (\ref{eq:diffR}) and noting that one can divide by $|\zeta|=1$ without any change, we are led to the following minimization problem
\begin{equation}\label{eq:inf1alt}
\widehat\delta^R_0(S_1^\inel)=\min_{g\in H^\infty} \Vert g-\widehat h\Vert_{L^\infty},
\end{equation}
where instead of $h(\zeta)$ defined in (\ref{eq:h}) we have now
\beq\label{eq:hhat}
\widehat h(\zeta)= \frac{F^R(\zeta)}{\zeta}- \frac{F(0)}{\zeta}-\frac{1}{\pi}\int_{z_{\thr}}^{1+\eta} \frac{\sigma(y)}{y(y-\zeta)} dy.
\eeq
\subsection{Solution by duality theorem}\label{sec:dual}
We 
shall apply a ``duality theorem" in functional  optimization theory
\cite{Duren}, which  replaces the original minimization problem~(\ref{eq:inf1})
by a maximization problem in a different functional space, denoted as the
``dual'' space. This will
allow us to obtain the solution by means of a numerical algorithm. For previous applications 
of this technique in particle physics, see the recent review \cite{Caprini:2019osi}.

We start  by introducing some standard notations \cite{Duren}. One denotes by  $H^p$, $p<\infty$, the class of functions $F(z)$ which are
analytic inside the unit disk $|z|<1$ and satisfy the boundary condition
\begin{equation}\label{eq:Hp}
\|F\|_{L^p}\equiv \left[\frac{1}{ 2\pi}\int_0^{2\pi}\vert F(e^{i\theta})\vert^p d\theta\right]^{1/p} <\infty.
\end{equation}
In particular, $H^1$ is the Banach space of analytic functions
with integrable modulus on the boundary, and $H^2$ is the Hilbert space
of analytic functions with integrable modulus squared. We introduced already
  the class  $H^\infty$ of functions analytic and bounded in
 $\vert z\vert <1$, for which the $L^\infty$ norm was defined
in (\ref{eq:diffR}). The classes $H^p$ and $H^q$ are said to be dual if the relation $1/p+1/q=1$ holds \cite{Duren}. It follows that $H^1$ and $H^\infty$ are dual to each other, while $H^2$ is dual to itself.

We now state the duality theorem of interest. 
Let the function  $h(\zeta=\mbox{exp}(i\theta))$ be an element of $L^\infty$.
Then the following equality holds (see Sect. 8.1 of Ref.~\cite{Duren}):
\begin{equation}\label{eq:dual}
\min_{g\in H^\infty}\|g-h\|_{L^\infty}=\sup_{G\in S^1}
\frac{1}{2\pi}\left\vert\oint_{\vert \zeta\vert =1} G(\zeta)h(\zeta) d \zeta
\right\vert,
\end{equation}
where $S^1$ denotes the unit sphere of the Banach space $H^1$, i.e. the set of functions 
$G\in H^1$ which satisfy the condition $ \|G\|_{L^1}\leq 1$.

We note first that the equality~(\ref{eq:dual}) is automatically satisfied  if $h(\zeta)$
 is the boundary value of an analytic 
function in the unit disk, since in this case
the minimal norm on the left-hand side is zero, and the right-hand side of Eq.~(\ref{eq:dual}) vanishes too, by Cauchy theorem.
 The nontrivial case corresponds  to a function $h(\zeta)$ 
 which  is not the
boundary value of a function analytic in
 $|z|< 1$. In that case, on the boundary $\zeta=\exp(i \theta)$, the function $h(\zeta)$ will admit the general Fourier expansion
\beq\label{eq:hpm}
h(\zeta)=\sum\limits_{n=0}^\infty h_n \zeta^n + \sum\limits_{n=1}^\infty c_n \zeta^{-n}\,, 
\eeq
where, due to reality property, the coefficients $h_n$ and $c_n$ are real. The analytic continuation of the expansion (\ref{eq:hpm})  inside $|z|<1$ will contain both an analytic part (the first sum) and a nonanalytic part (the second sum). Intuitively, we expect the minimum norm 
in (\ref{eq:dual}) to depend explicitly only on the nonanalytic part, i.e. on the coefficients $c_n$. The proof given below will confirm this expectation.

In order to evaluate the supremum on the right-hand side of (\ref{eq:dual}) we
 use  a factorization theorem (see the proof of Theorem~3.15 in Ref.~\cite{Duren}) according to which
every function $G(z)$ belonging to the unit sphere $S^1$ of $H^1$ 
 can be written as
\begin{equation}\label{eq:factor}
G(z)=w(z)f(z),
\end{equation}
where the functions $w(z)$ and $f(z)$ belong to the unit sphere $S^2$ of $H^2$,
i.e. are analytic and satisfy the conditions
\begin{equation}\label{wq:wGnorm}
\Vert w\Vert_{L^2} \leq 1\,,\quad \quad\quad\Vert f\Vert_{L^2}\leq 1\,.
\end{equation}
 Therefore, if one writes the Taylor expansions
 \begin{equation}\label{eq:wf}
w(z)=\sum_{n=0}^\infty w_nz^n,\quad \quad f(z)= \sum_{m=0}^\infty f_m z^m,
\end{equation}
the coefficients satisfy  the conditions
\begin{equation}\label{eq:wfl2}
\sum_{n=0}^\infty w_n^2\leq 1,\quad\quad \quad\sum_{m=0}^\infty f_m^2\leq 1.
\end{equation} 
After introducing the representation~(\ref{eq:factor}) in the r.h.s. of (\ref{eq:dual}), we obtain the equivalent  relation
\begin{equation}\label{eq:dual1}
\min_{g\in H^\infty}\|g-h\|_{L^\infty}=\sup_{w,f}\left\vert \frac{1}{2\pi}
\oint\limits_{\vert \zeta\vert =1} w(\zeta)f(\zeta)h(\zeta) d\zeta\right\vert,
\end{equation}
where the supremum on the right-hand side is taken with respect to the functions 
$w$ and $f$ with
the properties (\ref{eq:wf}) and (\ref{eq:wfl2}). Using in  (\ref{eq:dual1}) 
the expansions (\ref{eq:wf})
we obtain, after a straightforward calculation
\begin{equation}\label{eq:dual2}
\min_{g\in H^\infty}\|g-h\|_{L^\infty}=\sup_{\{w_n,f_m\}}
\left\vert\sum_{m,n=1}^\infty {\cal H}_{nm}w_{n-1}f_{m-1}\right\vert.
\end{equation}
Here the supremum is taken with respect to the sequences 
$w_n$ and $f_m$ subject to the condition (\ref{eq:wfl2}), and the numbers
\beq\label{eq:hank}
{\cal H}_{nm}=c_{n+m-1}, \quad n,m\geq 1,
\eeq
define a matrix ${\cal H}$ in terms of the coefficients 
\beq\label{eq:cn}
c_{n}=\frac{1}{2\pi i}\oint\limits_{\vert \zeta\vert=1}\zeta^{n-1}h(\zeta) d\zeta, \quad n\ge 1.
\eeq 
Matrices with elements defined as in (\ref{eq:hank}) are called Hankel matrices \cite{Duren}.

 Using the expression (\ref{eq:h}) of the function $h$, the coefficients (\ref{eq:cn}) can be written as
\be\label{eq:cn1}
c_n=\frac{1}{2\pi}\int\limits_0^{2 \pi} e^{i n\theta} F^R(e^{i \theta} )d\theta +\frac{1}{\pi}\int\limits_0^{1} y^{n-1} \sigma(y) dy,
\ee
where the quantities $F^R$ and $\sigma(y)$  are defined in the relations (\ref{eq:FR}) and \ref{eq:sigma}). Note that $c_n$ depend only on $\sigma(y)$ for $y<1$, because the contour integral in (\ref{eq:cn}) vanishes for $y>1$.

If we interpret $w_{n-1}$ and $\sum_m{\cal H}_{nm}f_{m-1}$ as the components of vectors
${\bf w}$ and ${\bf {\cal H}f}$,  the
absolute value of the sum in  (\ref{eq:dual2}) can be written as
$\left\vert{\bf w}\cdot{\bf {\cal H}f}\right\vert$, and the Cauchy--Schwarz inequality implies that it satisfies
\beq
\label{eq:CS}
\left\vert{\bf w}\cdot{\bf {\cal H}f}\right\vert\le\Vert{\bf w}\Vert_{L^2}
\Vert{\bf {\cal H}f}\Vert_{L^2}\le\Vert{\bf {\cal H}f}\Vert_{L^2}\,.
\eeq
Since (\ref{eq:CS}) is saturated for ${\bf w}\propto{\bf {\cal H}f}$, it follows that the
supremum in  (\ref{eq:dual2}) is given by the $L^2$ norm of ${\bf {\cal H}}$.
The solution of the minimization problem~(\ref{eq:inf1}) can therefore be written as 
\begin{equation}\label{eq:delta0}
\delta_0^R=\Vert{\cal H}\Vert_{L^2}=\Vert{\cal H}\Vert\,,
\end{equation}
where $\Vert{\cal H}\Vert$ is the spectral norm, given by 
 the square root of the greatest
 eigenvalue of the  matrix ${\cal H}^\dagger {\cal H}$.

In the numerical  calculations,  the matrix  ${\cal H}^\dagger {\cal H}$ has been truncated  at a finite order 
$N$, using the fact that for large $N$ the successive approximants approach the exact result (for a proof see appendix E of \cite{Ciulli:1973cg}).
  By the duality theorem, the initial
functional minimization problem (\ref{eq:inf1})
 was reduced to a rather simple numerical computation. 

It can be easily seen that the solution of the extremal problem (\ref{eq:inf1alt}) is obtained as the norm 
\begin{equation}\label{eq:delta0hat}
\widehat\delta_0^R(S_1)=\Vert\widehat{\cal H}\Vert,
\end{equation}
of the new Hankel matrix 
\beq\label{eq:hankhat}
\widehat{\cal H}_{nm}=\widehat c_{n+m-1}, \quad n,m\geq 1,
\eeq
defined in terms of the Fourier coefficients 
\bea\label{eq:cnhat}
\widehat c_n &=&\frac{1}{2\pi}\int\limits_0^{2 \pi} e^{i (n-1)\theta} F^R(e^{i \theta})d\theta 
+\frac{1}{\pi}\int\limits_0^{1} y^{n-2} \sigma(y) dy \nonumber\\
&-&\frac{S_1^\inel(Q^2)}{(2\nu_h)^\alpha}\delta_{n,1}.
\eea
We recall that $F^R(z)$ and $\sigma(x)$  are defined in the relations (\ref{eq:FR}) and \ref{eq:sigma}), and we used the relation (\ref{eq:F0}) connecting $F(0)$ to $S_1^\inel(Q^2)$.

We emphasize that the lower bounds calculated above are optimal. Moreover, as shown in \cite{Duren}, the minimum in the l.h.s. of the duality theorem (\ref{eq:dual}) is attained by an analytic function  with the required properties.



\begin{thebibliography}{99}

\bibitem{Cottingham:1963zz}
   W.~N.~Cottingham, The neutron proton mass difference and electron scattering experiments,
 \href{https://www.sciencedirect.com/science/article/abs/pii/000349166390023X?via%3Dihub}{Annals Phys.\  {\bf 25},  424 (1963)}

\bibitem{Harari:1966mu}
H.~Harari, Superconvergent  dispersion relations and electromagnetic mass differences, 
\href{https://journals.aps.org/prl/abstract/10.1103/PhysRevLett.17.1303}{Phys. Rev. Lett. \textbf{17}, 1303  (1966)}

\bibitem{Gross:1968zz}
D.~J.~Gross, H.~Pagels, Electromagnetic mass differences and Regge phenomenology,
\href{https://journals.aps.org/pr/abstract/10.1103/PhysRev.172.1381}{Phys. Rev. \textbf{172}, 1381  (1968)}


\bibitem{Creutz:1968ds}
M.~J.~Creutz, S.~D.~Drell,  E.~A.~Paschos,
High-energy limit for the real part of forward Compton scattering,
\href{https://journals.aps.org/pr/abstract/10.1103/PhysRev.178.2300}{Phys. Rev. \textbf{178},  2300 (1969)}

\bibitem{Damashek:1969xj}
M.~Damashek, F.~J.~Gilman, Forward Compton scattering,
\href{https://journals.aps.org/prd/abstract/10.1103/PhysRevD.1.1319}{Phys. Rev. D \textbf{1}, 1319  (1970)}

\bibitem{Dominguez:1970wu}
C.~A.~Dominguez, C.~Ferro Fontan,  R.~Suaya, Forward Compton scattering sum rules and j=0 singularities,
\href{https://www.sciencedirect.com/science/article/abs/pii/0370269370901978?via%3Dihub}{Phys. Lett. B \textbf{31}, 365 (1970)}

\bibitem{Elitzur:1970yy}
  M.~Elitzur,  H.~Harari,  Electromagnetic mass differences and inelastic electron scattering,
\href{https://www.sciencedirect.com/science/article/abs/pii/0003491670900060?via%3Dihub}{Annals Phys. \textbf{56}, 81  (1970)}.

\bibitem{Brodsky:1971zh}
S.J.~Brodsky, F.E.~Close, J. F.~Gunion, Compton scattering and fixed poles in parton field theoretical models, 
\href{https://journals.aps.org/prd/abstract/10.1103/PhysRevD.5.1384}{Phys. Rev. D \textbf{5}, 1384  (1972)}

 \bibitem{Zee:1971df}
A.~Zee, The proton-neutron mass difference problem and related topics,
\href{https://www.sciencedirect.com/science/article/abs/pii/0370157372900075?via%3Dihub}{Phys. Rept. \textbf{3}, 127 (1972)}

\bibitem{Brodsky:1972vv}
S.J.~Brodsky, F.E.~Close, J. F.~Gunion, Phenomenology of photon processes, vector dominance and crucial tests for parton models,
\href{https://journals.aps.org/prd/abstract/10.1103/PhysRevD.6.177}{Phys. Rev. D \textbf{6}, 177 (1972)}

 \bibitem{Brodsky:2008qu}
  S.J. Brodsky, F.J. Llanes-Estrada,  A.P. Szczepaniak, Local two-photon couplings and the $J=0$ fixed pole in real and virtual Compton scattering,
 \href{https://journals.aps.org/prd/abstract/10.1103/PhysRevD.79.033012}{Phys.\ Rev.\ D {\bf 79}, 033012  (2009)}, 
 \href{https://arxiv.org/abs/0812.0395}{arXiv:0812.0395}

\bibitem{Muller:2015vha}
  D. M\"uller,  K.M. Semenov-Tian-Shansky,  $J=0$ fixed pole and $D$-term form factor in deeply virtual Compton scattering, \href{https://journals.aps.org/prd/abstract/10.1103/PhysRevD.92.074025}{ Phys. Rev. D {\bf 92},  074025 (2015)}, 
  \href{https://arxiv.org/abs/1507.02164}{arXiv:1507.02164}

\bibitem{Gasser:1974wd}
J.~Gasser, H.~Leutwyler, Implications of scaling for the proton-neutron mass  difference,
\href{https://www.sciencedirect.com/science/article/abs/pii/0550321375904939?via%3Dihub}{Nucl. Phys. B \textbf{94},  269 (1975)}

\bibitem{WalkerLoud:2012bg} 
  A.~Walker-Loud, C.~E.~Carlson, G.~A.~Miller, The electromagnetic self-energy contribution to $M_p - M_n$ and the isovector nucleon magnetic polarizability,
\href{https://journals.aps.org/prl/abstract/10.1103/PhysRevLett.108.232301}{Phys.\ Rev.\ Lett.\  {\bf 108},  232301 (2012)}, 
 \href{https://arxiv.org/abs/1203.0254}{arXiv:1203.0254}

\bibitem{WalkerLoud:2012en}
A.~Walker-Loud, C.~E.~Carlson, G.~A.~Miller, Cottingham formula for the electromagnetic self-energy contribution to $M_p - M_n$,
\href{https://pos.sissa.it/164/136}{PoS LATTICE2012, 136 (2012) }, \href{https://arxiv.org/abs/1210.7777}{arXiv:1210.7777}.

\bibitem{Erben:2014hza}
  F.B. Erben, P. E. Shanahan, A.W. Thomas,  R. D. Young, Dispersive estimate of the electromagnetic charge symmetry violation in the octet baryon masses,
\href{https://journals.aps.org/prc/abstract/10.1103/PhysRevC.90.065205}{ Phys.\ Rev.\ C {\bf 90},  065205 (2014)}, 
 \href{https://arxiv.org/abs/1408.6628}{arXiv:1408.6628} 

\bibitem{Thomas:2014dxa}
A.~W.~Thomas, X.~G.~Wang,  R.~D.~Young, Electromagnetic contribution to the proton-neutron mass splitting,
\href{https://journals.aps.org/prc/abstract/10.1103/PhysRevC.91.015209}{Phys. Rev. C \textbf{91} (2015) 015209}, 
\href{https://arxiv.org/abs/1406.4579}{arXiv:1406.4579}

\bibitem{Gasser:2015dwa}
J.~Gasser, M.~Hoferichter, H.~Leutwyler, A.~Rusetsky, Cottingham formula and nucleon polarisabilities,
\href{https://link.springer.com/article/10.1140%2Fepjc%2Fs10052-015-3580-9}{Eur. Phys. J. C \textbf{75},  375 (2015)}; Erratum:\href{https://link.springer.com/article/10.1140%2Fepjc%2Fs10052-020-7905-y}{Eur. Phys. J. C \textbf{80}, 353 (2020)}, \href{https://arxiv.org/abs/1506.06747}{arXiv:1506.06747}

\bibitem{Caprini:2016wvy}
I.~Caprini, Constraints on the virtual Compton scattering on the nucleon in a new dispersive formalism,
\href{https://journals.aps.org/prd/abstract/10.1103/PhysRevD.93.076002}{Phys. Rev. D \textbf{93}, 076002  (2016)}, \href{https://arxiv.org/abs/1601.02787}{arXiv:1601.02787}

\bibitem{Walker-Loud2018}
   A.~Walker-Loud, On the Cottingham formula and the electromagnetic contribution to the proton-neutron mass splitting,
 \href{https://pos.sissa.it/317/045}{PoS CD {\bf 2018}, 045 (2019)},
 \href{https://arxiv.org/abs/1907.05459}{arXiv:1907.05459}


\bibitem{Tomalak:2018dho}
O.~Tomalak, Electromagnetic proton-neutron mass difference,
\href{https://link.springer.com/article/10.1140%2Fepjp%2Fs13360-020-00413-9}{Eur. Phys. J. Plus \textbf{135},  411 (2020)}, 
\href{ https://arxiv.org/abs/1810.02502}{arXiv:1810.02502}


\bibitem{Gasser:2020hzn}
J.~Gasser, H.~Leutwyler, A.~Rusetsky, Sum rule for the Compton amplitude and implications for the proton-neutron mass difference,
\href{https://link.springer.com/article/10.1140%2Fepjc%2Fs10052-020-08615-2}{Eur. Phys. J. C \textbf{80}, 1121  (2020)}, 
\href{https://arxiv.org/abs/2008.05806}{arXiv:2008.05806}

\bibitem{Gasser:2020mzy}
J.~Gasser, H.~Leutwyler,  A.~Rusetsky, On the mass difference between proton and neutron, 
\href{https://www.sciencedirect.com/science/article/pii/S0370269321000277?via%3Dihub}{Phys. Lett. B \textbf{814}, 136087  (2021)}, 
\href{https://arxiv.org/abs/2003.13612}{arXiv:2003.13612}

\bibitem{Caprini:2019osi}
  I.~Caprini, \textit{Functional analysis and optimization methods in hadron physics}, \href{https://www.springer.com/gp/book/9783030189471}{Book series: SpringerBriefs in Physics (2019)}

\bibitem{Duren} P. Duren, \textit{Theory of $H^p$ Spaces}, New York, Academic (1970)


\bibitem{Drechsel:2007if}
  D.~Drechsel, S.~S.~Kamalov,  L.~Tiator, Unitary isobar model - MAID2007,
  \href{https://link.springer.com/article/10.1140%2Fepja%2Fi2007-10490-6}{Eur.\ Phys.\ J.\ A {\bf 34},  69  (2007)}, 
  \href{https://arxiv.org/abs/0710.0306}{arXiv:0710.0306}
  
\bibitem{Kamalov:2000en}
  S.~S.~Kamalov, S.~N.~Yang, D.~Drechsel, O.~Hanstein,  L.~Tiator, $\gamma^* N \to \Delta$ transition form-factors: a new analysis of the JLab data on $p (e, e' p) \pi^0$ at $Q^2=$2.8 and 4.0 $(\gev/c)^2$, 
 \href{https://journals.aps.org/prc/abstract/10.1103/PhysRevC.64.032201}{Phys.\ Rev.\ C {\bf 64},  032201  (2001)}, 
 \href{https://arxiv.org/abs/nucl-th/0006068} {arXiv:nucl-th/0006068}
 
\bibitem{Hilt:2013fda}
  M.~Hilt, B.~C.~Lehnhart, S.~Scherer, L.~Tiator, Pion photo- and electroproduction in relativistic baryon chiral perturbation theory and the chiral MAID interface,
 \href{https://journals.aps.org/prc/abstract/10.1103/PhysRevC.88.055207}{Phys.\ Rev.\ C {\bf 88},  055207 (2013)}, 
\href{https://arxiv.org/abs/1309.3385}{arXiv:1309.3385}

\bibitem{MAID} \href{http://portal.kph.uni-mainz.de/MAID/}{http://portal.kph.uni-mainz.de/MAID/}


\bibitem{Bosted:2007xd}
  P.~E.~Bosted,  M.~E.~Christy, Empirical fit to inelastic electron-deuteron and electron-neutron resonance region transverse cross-sections,
 \href{https://journals.aps.org/prc/abstract/10.1103/PhysRevC.77.065206}{Phys.\ Rev.\ C {\bf 77} (2008) 065206}, 
\href{https://arxiv.org/abs/0711.0159} {arXiv:0711.0159}
 
\bibitem{Christy:2007ve}
  M.~E.~Christy, P.~E.~Bosted, Empirical fit to precision inclusive electron-proton cross- sections in the resonance region,
\href{https://journals.aps.org/prc/abstract/10.1103/PhysRevC.81.055213 }{Phys.\ Rev.\ C {\bf 81},  055213   (2010)},
 \href{https://arxiv.org/abs/0712.3731}{arXiv:0712.3731}
 
\bibitem{Alwall:2004wk}
  J.~Alwall and G.~Ingelman, Interpretation of electron-proton scattering at low $Q^2$,
  \href{https://www.sciencedirect.com/science/article/pii/S0370269304009554?via%3Dihub}{Phys.\ Lett.\ B {\bf 596}, 77   (2004)}
  \href{https://arxiv.org/abs/hep-ph/0402248}{hep-ph/0402248}

\bibitem{Alekhin:2013nda}
 S.~Alekhin, J.~Bl\"umlein,  S.~Moch, The ABM parton distributions tuned to LHC data,
 \href{https://journals.aps.org/prd/abstract/10.1103/PhysRevD.89.054028}{Phys.\ Rev.\ D {\bf 89}, 054028  (2014)}, 
 \href{https://arxiv.org/abs/1310.3059}{arXiv:1310.3059}

\bibitem{Alekhin:2017kpj}
  S.~Alekhin, J.~Bl\"umlein, S.~Moch, R.~Placakyte, Parton distribution functions, $\alpha_s$, and heavy-quark masses for LHC Run II,
 \href{https://journals.aps.org/prd/abstract/10.1103/PhysRevD.96.014011}{Phys.\ Rev.\ D {\bf 96}, 014011  (2017)}, \href{ https://arxiv.org/abs/1701.05838}{arXiv:1701.05838}

\bibitem{Alekhin:2019ntu}
 S.~Alekhin, J.~Bl\"umlein,  S.~Moch, An update of the ABM16 PDF fit, 
 \href{https://pos.sissa.it/352/002}{PoS DIS {\bf 2019}, 002 (2019)},  \href{https://arxiv.org/abs/1909.03533}{arXiv:1909.03533} 

\bibitem{Ciulli:1973cg}
S.~Ciulli, G.~Nenciu, Optimal analytic extrapolation for the scattering amplitude from cuts to interior points,
\href{https://aip.scitation.org/doi/10.1063/1.1666242}{J. Math. Phys. \textbf{14}, 1675 (1973)}

\end{thebibliography}
\end{document}